\documentclass{aastex7}

\usepackage{enumitem}
\usepackage{graphicx}
\usepackage{subfig}

\accepted{21 November, 2025}

\submitjournal{RNAAS}

\begin{document}

\title{A New FU Orionis Accretion Outburst in the W5 HII Region}

\author{Lynne A. Hillenbrand}
\affiliation{Department of Astronomy, California Institute of Technology Pasadena, CA, 91125, USA}
\email{lah@astro.caltech.edu}

\author{Matthew J. Graham} 
\affiliation{Department of Astronomy, California Institute of Technology Pasadena, CA, 91125, USA}
\email{mjg@caltech.edu}

\author{Mansi M. Kasliwal} 
\affiliation{Department of Astronomy, California Institute of Technology Pasadena, CA, 91125, USA}
\email{mansi@astro.caltech.edu}

\author{Josiah Purdum} 
\affiliation{Department of Astronomy, California Institute of Technology Pasadena, CA, 91125, USA}
\email{jpurdum@caltech.edu}

\author{Jesper Sollerman}  
\affiliation{The Oskar Klein Centre, Department of Astronomy, AlbaNova, SE-106 91 Stockholm , Sweden} 
\email{jesper@astro.su.se}

\author[0000-0002-9540-853X]{Adolfo S. Carvalho}
\affiliation{Center for Astrophysics, Harvard University, 60 Garden St., Cambridge, MA, 02138, USA}
\email{adolfo.carvalho@cfa.harvard.edu}

\author[0000-0002-0631-7514]{Michael A. Kuhn}
\affiliation{School of Physics, Engineering \& Computer Science, University of Hertfordshire, College Lane, Hatfield, AL10 9AB, United Kingdom }
\email{m.kuhn@herts.ac.uk}

\author[0000-0001-7062-9726]{Roger Smith} 
\affiliation{Caltech Optical Observatories, California Institute of Technology, Pasadena, CA  91125, USA}
\email{rsmith@astro.caltech.edu}

\author[0000-0003-1412-2028]{Michael C.~B.\ Ashley} 
\affiliation{School of Physics, University of New South Wales, 2052, Australia}
\email{m.ashley@unsw.edu.au}
\author{Nicholas Earley}  
\affiliation{Department of Astronomy, California Institute of Technology Pasadena, CA, 91125, USA}
\email{nearley@caltech.edu}

\author[0000-0003-2451-5482]{Russ R. Laher} 
\affiliation{IPAC, California Institute of Technology, Pasadena, CA 91125, USA}
\email{xchen@ipac.caltech.edu}

\author[0000-0001-9152-6224]{Tracy X. Chen} 
\affiliation{IPAC, California Institute of Technology, Pasadena, CA 91125, USA}
\email{laher@ipac.caltech.edu}

\begin{abstract}
We announce a recently detected outburst that is currently only a few months old, and probably of FU Orionis type. 
The progenitor to the outburst was an emission-line, flat-spectrum SED 
young stellar object located in the W5 region, 
though somewhat outside the main star formation action.
We present optical, near-infrared, and mid-infrared lightcurves 
that illustrate the quiescent state of [KAG2008] 13656 and its subsequent 
$\Delta r \approx -4$ mag and $\Delta J\approx -3$ mag outburst 
over $\sim$75 days in late-2025.
Follow-up optical and near-infrared spectroscopy confirms the expected features from an FU Ori disk and outflow.  
\end{abstract}


\section{Introduction } \label{sec:intro}

Episodic accretion events in young stellar objects (YSOs) take many forms, with the longest duration and
largest amplitude outbursts known as FU Ori stars - named after the prototype FU Ori \citep{Herbig1966}.
These outbursts are characterized by extreme brightening that can be $\sim4-6$ mag in the optical and $\sim 1-3$ mag in the infrared,
and have rise times of a few months to a few years.  The outbursts are the result of disk instabilities
and accretion rate increases \citep{Hartmann1996} that drive the accretion luminosity upward by
factors of typically hundreds.
In an FU Ori outburst state, optical and near-infrared observations are dominated
by the accretion disk atmosphere.

Understanding the processes of stellar mass assembly, through disk accretion, 
requires understanding the rapid mass accumulation that occurs
during FU Ori events \citep{Fischer2023}.  
Understanding the evolution of circumstellar disks into planets 
must also account for the effects of significantly variable heating, 
especially from ultraviolet photons, on disk dust and gas chemistry \citep{Molyarova2018}.

W5 is a large and multi-bubble HII region in the outer Galaxy, with distance $2.2-2.4$ kpc \citep{Damian2021},
also known as IC 1848.  There are several massive star clusters with age $\sim2$ Myr \citep{Damian2024}
each harboring multiple O-type stars.

Our source of interest at RA=02:58:07.83, Decl. = +61:12:57.09 (J2000.)
is just north of W5/IC 1848 East.
It has a well-characterized pre-outburst spectral energy distribution from the optical to the mid-infrared (Figure~\ref{fig}b).
The source was cataloged as a Class II disk in the Spitzer study of the W5 region by \cite{Koenig2008},
and similarly tagged by \cite{Marton2016} and \cite{Winston2020}. 
It also appears in IPHAS as an H$\alpha$ emission-line object.
The location, SED, and evidence for emission lines all indicate [KAG2008] 13656 as a secure YSO.

\section{Photometric Outburst Detection and Spectroscopic Follow-up} \label{sec:middle}

\begin{figure}[ht]
    \centering
\subfloat[]{\includegraphics[width=0.38\linewidth]{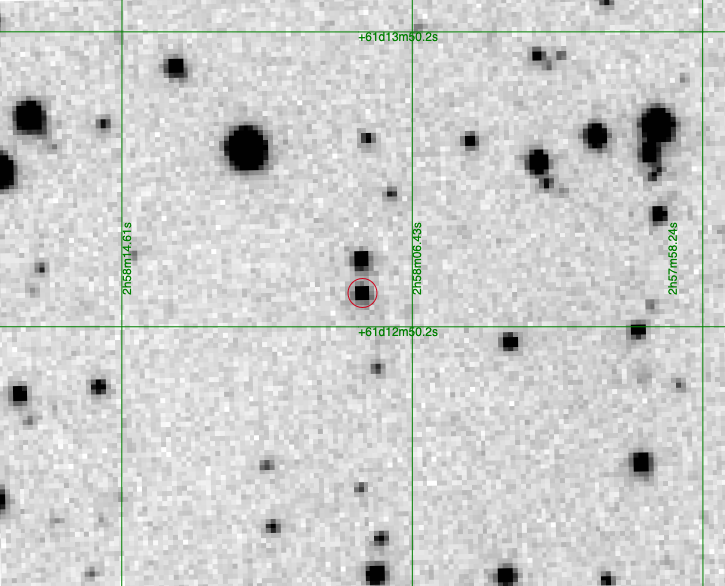}}
\hfill
\subfloat[]{\raisebox{2.5truein}{\includegraphics[width=0.38\linewidth, angle=-90, trim=0in 0in 1.75in 0in, clip]{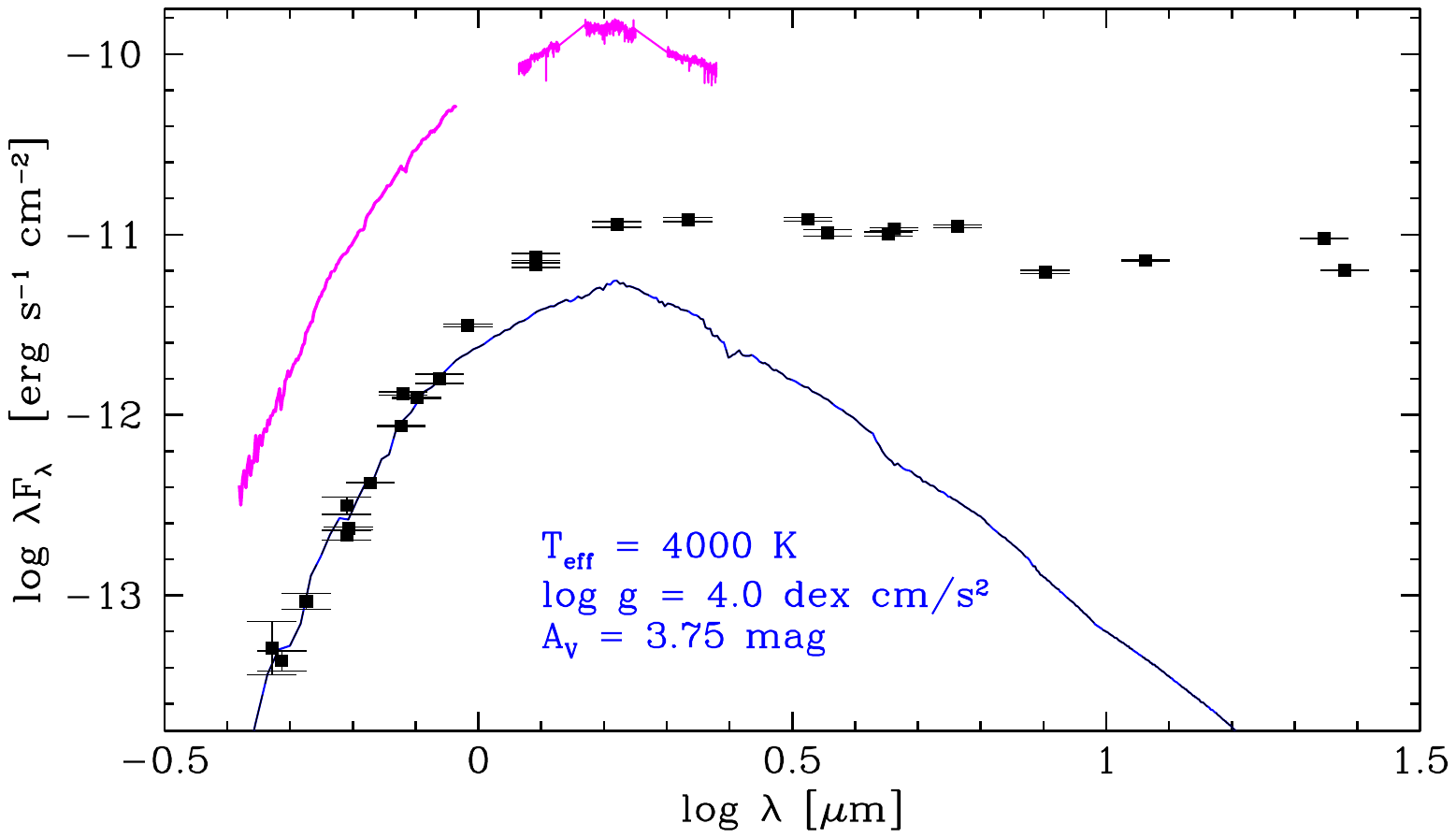}}}
\\ [-1em]
\subfloat[]{\raisebox{0.0truein}{\includegraphics[width=0.55\linewidth]{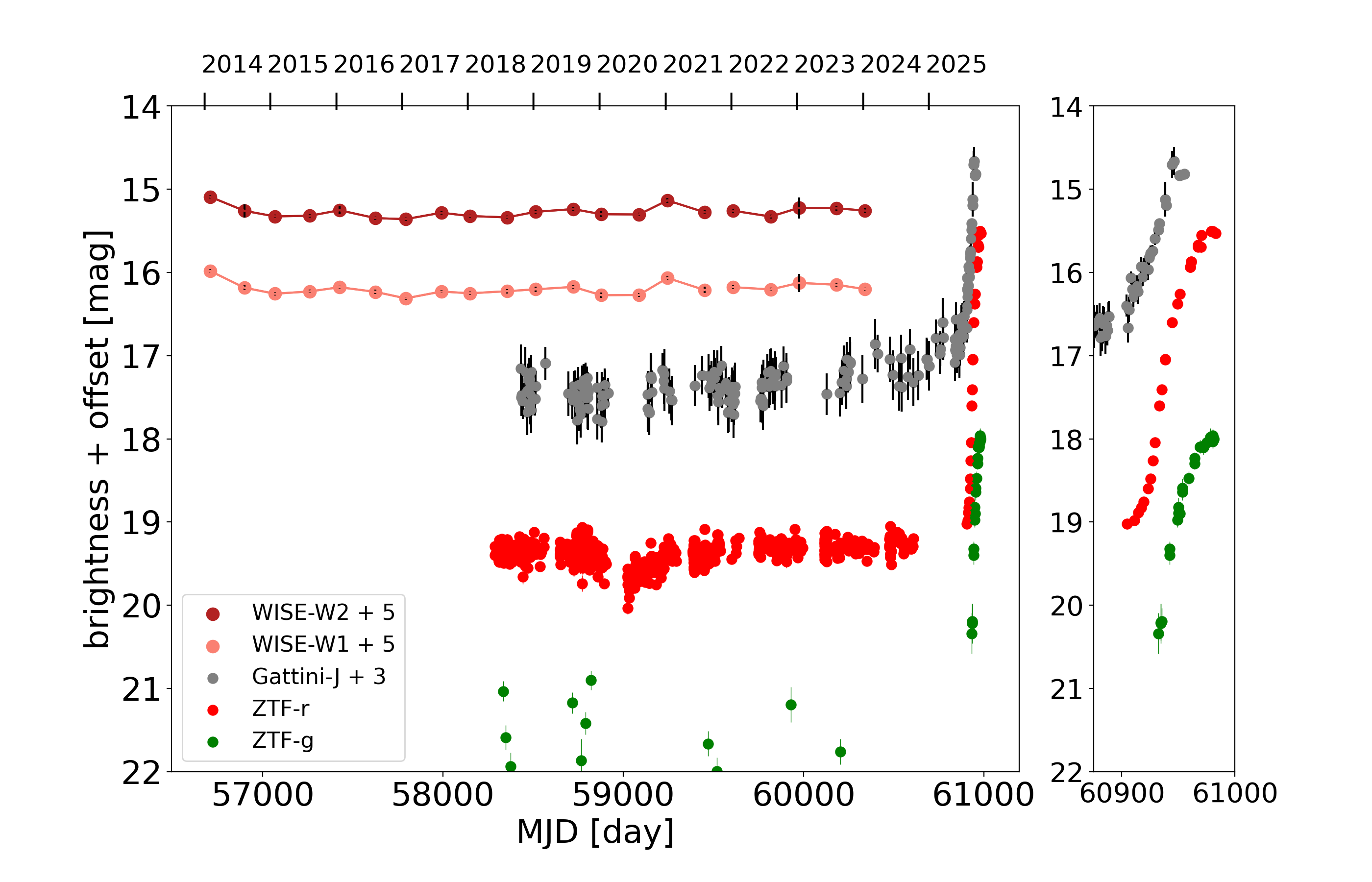}}}
\hfill
\subfloat[]{\raisebox{+0.1truein}{\includegraphics[width=0.45\linewidth]{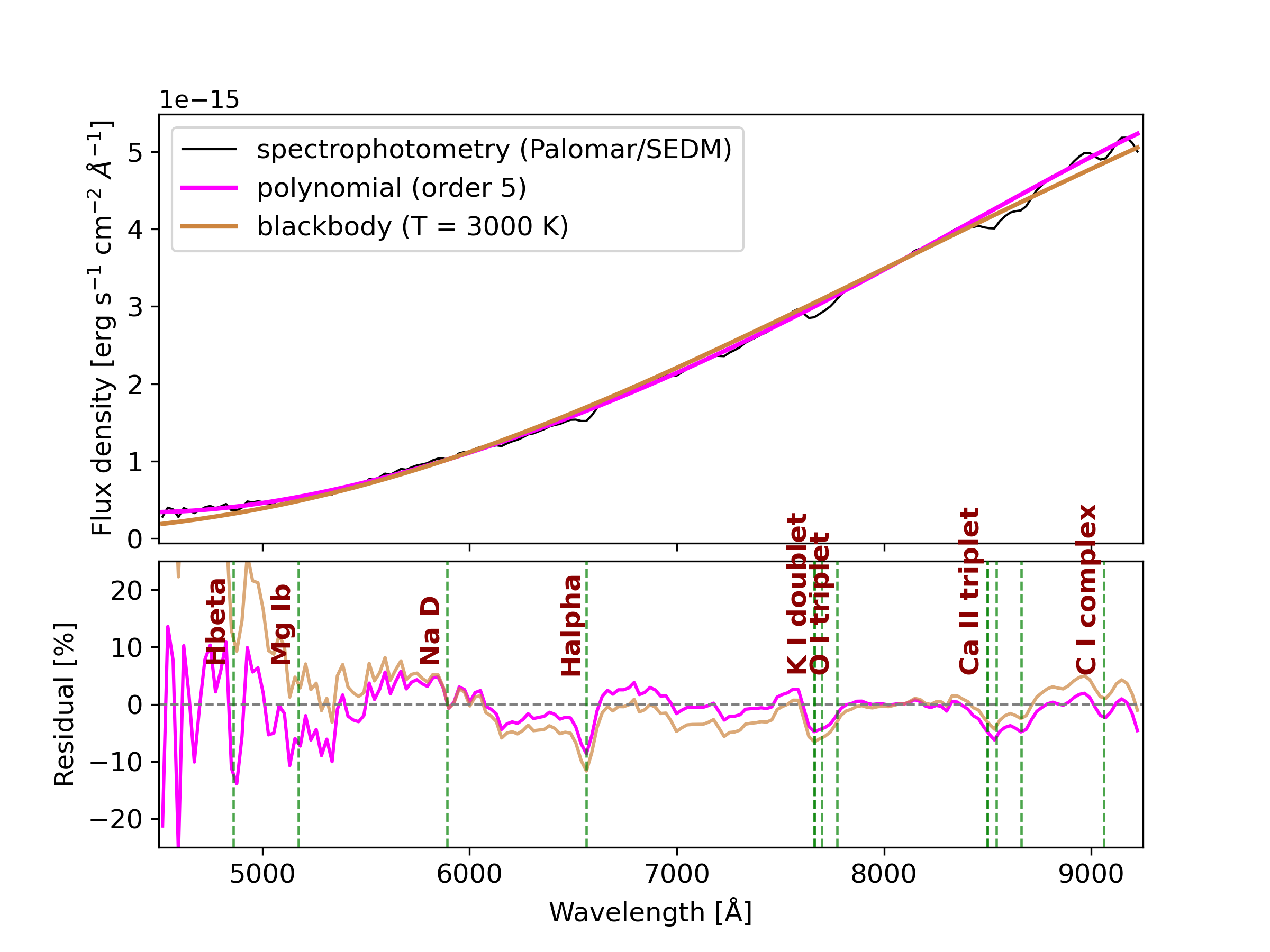}}}
    \caption{\textbf{Top left:} 
    ZTF r-band image (N/E are top/left) of the pre-outburst source and environment.
    \textbf{Top right:} 
    SEDs assembled for the progenitor source (black points) along with a 
    plausible stellar model (blue line), and for the outbursting source (magenta).
    \textbf{Bottom left:} 
    Lightcurves in optical ZTF g-band (0.48 $\mu$m), r-band (0.64 $\mu$m); 
    near-infrared Gattini J-band (1.6 $\mu$m); 
    and mid-infrared NEOWISE W1 (3.6 $\mu$m), W2 (4.5 $\mu$m) bands.
    \textbf{Bottom right:} Optical spectrophotometry highlighting absorption in plausible disk/wind lines. 
    }
    \label{fig}
\end{figure}

Initial detection of the source brightening was due to a specially designed filter running on the 
Zwicky Transient Facility \citep[ZTF;][]{Bellm2019} alerts stream \citep{Patterson2019}.
The filter selects only certain ZTF fields containing large star forming regions,
and identifies sources with sustained brightness increases that are
within the mag/month range exhibited by previous FU Ori outbursts.
Although previously detected 
as a variable object and earning the name ZTF20abjgfwm, 
around MJD=60905 (2025-08-18 UT), the source began a 
seemingly colorless brightening event 
and passed our alert filter $\sim40$ days later.
The source is also detected in the Gattini time domain survey \citep{De2020}.
Figure~\ref{fig}c shows the lightcurve assembled to date, 
including from NEOWISE \citep{Mainzer2014}
which unfortunately stops prior to the outburst.
Lightcurve peak appears to be around MJD=60980.

The source was subsequently observed with the Palomar 60" telescope's SED Machine \citep[SEDM;][]{Blagorodnova2018} on 2025-11-01 UT.
Figure~\ref{fig}d shows the steeply rising continuum to the red,
and clear indication of shallow absorption in several key features: H$\alpha$, H$\beta$, \ion{Na}{1}D, 
and the \ion{K}{1}/doublet $+$ \ion{O}{1}/triplet that are blended at the $R\approx100$ resolution.
These lines are typical of the strong outflows seen in FU Ori stars.  
No emission is present, though prior to outburst H$\alpha$ was in emission.

The KeckI telescope and MOSFIRE \citep{McLean2012} were used on 2025-11-12 UT 
to obtain the $R\approx3500$
infrared spectrum in Figure~\ref{fig}b, which exhibits the 
mix of molecular ($H_2O$, $CO$) and atomic (\ion{Na}{1}, \ion{Ca}{1}, \ion{Al}{1}, \ion{Mg}{1}, \ion{Si}{1})  
features expected for a solid FU Ori classification.

\section{Discussion}

Should the current photometric and spectroscopic source characteristics persist,
following the convention introduced in \cite{hillenbrand2025},
the source would be named FUOr-Cas 0258+6112.
Multi-wavelength follow-up is encouraged, 
especially over the next few months to year, as the spectrum and lightcurves likely evolve.

\bibliography{ms}{}
\bibliographystyle{aasjournal}

\end{document}